\documentclass[twocolumn,showpacs,prb]{revtex4-1}%
\usepackage{amsfonts}
\usepackage{amsmath}
\usepackage{amssymb}
\usepackage{graphicx}
\usepackage{txfonts}%
\providecommand{\U}[1]{\protect\rule{.1in}{.1in}}

\begin{document}
\title{Selective generation and amplification of RKKY interactions by P-N interface}
\author{Shu-Hui Zhang$^{1,2}$}
\author{Jia-Ji Zhu$^{3}$}
\author{Wen Yang$^{1}$}
\email{wenyang@csrc.ac.cn}
\author{Kai Chang$^{4,5}$}
\email{kchang@semi.ac.cn}
\affiliation{$^{1}$Beijing Computational Science Research Center, Beijing 100193, China}
\affiliation{$^{2}$College of Science, Beijing University of Chemical Technology, Beijing 100029, China}
\affiliation{$^{3}$Institute for quantum information and spintronics, School of Science,
Chongqing University of Posts and Telecommunications, Chongqing 400065, China}
\affiliation{$^{4}$SKLSM, Institute of Semiconductors, Chinese Academy of Sciences, P.O.
Box 912, Beijing 100083, China}
\affiliation{$^{5}$Synergetic Innovation Center of Quantum Information and Quantum Physics,
University of Science and Technology of China, Hefei, Anhui 230026, China}

\begin{abstract}
We propose a physical mechanism to generate and selectively amplify
anisotropic Rudermann-Kittel-Kasuya-Yosida (RKKY) interactions between two
local spins. The idea is to combine the deflection of the carrier velocity by
a P-N interface and the locking of this velocity to the carrier spin
orientation via spin-orbit coupling. We provide analytical and numerical
results to demonstrate this mechanism on the surface of a topological
insulator P-N junction. This work identifies the P-N interface as a second
knob which, together with the carrier density, enables independent control
of the strength and anisotropy of the RKKY\ interaction over a wide range.
These findings may be relevant to scalable quantum computation and
two-impurity quantum criticality.

\end{abstract}

\maketitle

\section{Introduction}

The carrier-mediated Rudermann-Kittel-Kasuya-Yosida (RKKY) interaction
\cite{RudermanPR1954,KasuyaPTP1956,YosidaPR1957,DietlRMP2014} between local
spins has been known for more than sixty years as an important mechanism for
the broadening of nuclear spin resonance lines, magnetic ordering, and spin
glass behaviors in magnetic metals, alloys, and semiconductors. It
characterizes the magnetic response of the itinerant carriers and finds
interesting applications, e.g., it enables entanglement between spatially
separated spins in quantum dots
\cite{PiermarocchiPRL2002,CraigScience2004,RikitakePRB2005,FriesenPRL2007,SrinivasaPRL2015}
for scalable quantum computation \cite{LossPRA1998,TrifunovicPRX2012} and
allows magnetic ordering of magnetically doped materials for spintronics
\cite{WolfScience2001,ZuticRMP2004,PesinNatMater2012} and new topological
phases \cite{LiuPRL2013a,WangPRL2014,KlinovajaPRL2013}. For a pair of magnetic
atoms, it competes with the Kondo effect and gives rise to rich physics \cite{JayaprakashPRL1981,JonesPRL1987,JonesPRL1988} that
have been of considerable theoretical and experimental interest for decades
\cite{GegenwartNatPhys2008,NeelPRL2011,PruserNatComm2014,SpinelliNatComm2015}. For these
applications, \textit{selective} amplification of different interaction terms
(beyond the isotropic Heisenberg-like term) is highly desirable.

Recently, the RKKY interaction was studied in many
materials, e.g., dilute ferromagnetic semiconductors (see Ref. \onlinecite{DietlRMP2014} for a review), graphene
\cite{VozmedianoPRB2005,DugaevPRB2006,BreyPRL2007,SaremiPRB2007,HwangPRL2008,UchoaPRL2011}%
, Dirac and Weyl semimetals \cite{ChangPRB2015,HosseiniPRB2015}, and the
surface of topological insulators (TIs)
\cite{LiuPRL2009,BiswasPRB2010,ZhuPRL2011,AbaninPRL2011,EfimkinPRB2014,LitvinovPRB2014,ZyuzinPRB2014}%
. The successful measurement \cite{WahlPRL2007,MeierScience2008,ZhouNatPhys2010,KhajetooriansNatPhys2012} of the RKKY interaction with atomic-scale spatial resolution \cite{ZhouNatPhys2010,KhajetooriansNatPhys2012} opens a new playground for engineering this interaction. Interesting theoretical and experimental works for amplifying the isotropic
Heisenberg-like term have been reported
\cite{BrovkoPRL2008,InosovPRL2009,XiuNatMater2010,Black-SchafferPRB2010,NeelPRL2011,PowerPRB2012,PowerPRB2012a,ChenPRB2013,YaoPRL2014,BouhassouneNatComm2014,KlierPRB2015,NieNatComm2016,Zhang2DMater2017}%
, but selective generation and amplification of anisotropic terms remains
elusive. The general consensus is that these terms originate from the
spin-orbit coupling of the carriers, so tuning these terms requires tuning the
symmetry of the spin-orbit coupling (e.g., combining the Rashba
and Dresselhaus terms \cite{ZhuPRB2010} or the warping effect of TI
surface states \cite{ZhuPRL2011}). The latter is determined by the crystal
symmetry and sample orientation, so it has no (or very limited) tunability.

Here we propose a physical mechanism to generate and selectively amplify
anisotropic interaction terms between two local spins $\hat{\mathbf{S}}_{1}$
and $\hat{\mathbf{S}}_{2}$. The idea is that these terms arise from
the spin orientation of carriers traveling between $\hat{\mathbf{S}}_{1}$ and
$\hat{\mathbf{S}}_{2}$, so the P-N interface can deflect the velocity and
hence the spin (via spin-orbit coupling) of those carriers to achieve
controlled generation of anisotropic terms. The negative refraction of
carriers across the P-N interface
\cite{CheianovScience2007,ZhaoPRL2013,LeeNatPhys2015,ChenScience2016} further
allows selective amplification of these terms. This physical mechanism is
applicable to the P-N junction of any material with spin-orbit coupling. For
specificity, here we demonstrate this mechanism by both analytical and
numerical results on the surface of a TI P-N junction, which forms the basis
of several interesting proposals
\cite{WangPRB2012,ZhaoPRL2013,HabibPRL2015,IlanPRL2015} and was recently
fabricated
\cite{LiACSNano2015,BathonAdvMater2016,TuNatComm2016,KimACSNano2017} with
atomically abrupt interfaces \cite{KimACSNano2017}.
Ever since the discovery of the RKKY interaction, extensive efforts have been devoted to this interaction, the vast majority of which focus on its manipulation in uniform materials via the carrier density. Here, our work identifies the P-N interface as a second knob which, together with the carrier density, enables
\textit{independent} control of the strength and anisotropy of the RKKY
interaction. These findings mark a step towards engineering RKKY\ interactions
in spin-orbit coupled systems for the nonlocal control of spin and entanglement.

This paper is organized as follows. In Sec. II, we give an intuitive physical
picture for the origin of the anisotropic terms. In Sec. III, we provide
analytical and numerical results for engineering the RKKY\ interactions on the
surface of TI P-N\ junctions. Finally we summarize our findings in Sec. IV.

\section{Physical picture}

We consider two local spins $\hat{\mathbf{S}}_{1}$ (located at $\mathbf{R}%
_{1}$) and $\hat{\mathbf{S}}_{2}$ (located at $\mathbf{R}_{2}$) coupled to
itinerant carriers via the contact exchange interaction $-\lambda\sum_{i}%
\hat{\mathbf{S}}_{i}\cdot\hat{\boldsymbol{\sigma}}\delta(\hat{\mathbf{r}%
}-\mathbf{R}_{i})$, where $\hat{\boldsymbol{\sigma}}$ are Pauli matrices for
the carrier spin. At low temperature, the carriers mediate an effective
interaction \cite{RudermanPR1954,KasuyaPTP1956,YosidaPR1957,BrunoPRB1995}
between $\hat{\mathbf{S}}_{1}$ and $\hat{\mathbf{S}}_{2}$:
\begin{equation}
\hat{H}_{\mathrm{RKKY}}=-\frac{\lambda^{2}}{\pi}\operatorname{Im}\int
_{-\infty}^{E_{F}}\hat{K}(E)dE,\label{RKKY_DEF}%
\end{equation}
where $E_{F}$ is the Fermi energy, $\operatorname{Im}\hat{O}\equiv(\hat
{O}-\hat{O}^{\dagger})/(2i)$ ($\forall\hat{O}$), and
\begin{equation}
\hat{K}(E)=\operatorname*{Tr}[\hat{\boldsymbol{\sigma}}\cdot\hat{\mathbf{S}%
}_{1}\hat{G}(\mathbf{R}_{1},\mathbf{R}_{2};E)\hat{\boldsymbol{\sigma}}%
\cdot\hat{\mathbf{S}}_{2}\hat{G}(\mathbf{R}_{2},\mathbf{R}_{1}%
;E)].\label{IE_DEF}%
\end{equation}
Here the carrier propagator (retarded Green's function) $\hat{G}%
(\mathbf{r}_{2},{\mathbf{r}}_{1};E)$ in real space is still an operator acting
on the carrier spin and the trace is taken over the carrier spin.

Equation (\ref{IE_DEF}) shows that the RKKY interaction arises from two steps:
(i)\ The carrier travels from $\hat{\mathbf{S}}_{1}$ to $\hat{\mathbf{S}}_{2}%
$, as described by $\hat{G}(\mathbf{R}_{2},\mathbf{R}_{1};E)$, and undergoes
the interaction $\hat{\boldsymbol{\sigma}}\cdot\hat{\mathbf{S}}_{2}$; (ii) The
carrier travels from $\hat{\mathbf{S}}_{2}$ back to $\hat{\mathbf{S}}_{1}$, as
described by $\hat{G}(\mathbf{R}_{1},\mathbf{R}_{2};E)$, and undergoes the
interaction $\hat{\boldsymbol{\sigma}}\cdot\hat{\mathbf{S}}_{1}$, which
restores the initial spin state of the carrier. This process mediates an
effective interaction $\sim\hat{K}(E)$ between $\hat{\mathbf{S}}_{1}$ and
$\hat{\mathbf{S}}_{2}$. The sum of the contributions from all the occupied
carrier states gives the total RKKY\ interaction in Eq. (\ref{RKKY_DEF}). When
the distance $R\equiv|\mathbf{R}_{2}-\mathbf{R}_{1}|$ exceeds the Fermi
wavelength of the carriers, stationary phase approximation
\cite{RothPR1966,LiuPRB2012a} shows that the energy integral in Eq.
(\ref{RKKY_DEF}) is dominated by the contribution near the Fermi level
\cite{PowerPRB2011,PowerPRB2012}: $\hat{H}_{\mathrm{RKKY}}\propto\hat{K}%
(E_{F})/R$, so the strength (anisotropy) of the RKKY\ interaction is
determined by the magnitude (spin polarization) of the carrier propagator on
the Fermi level.

In a $d$-dimensional uniform system, the propagators $\hat{G}(\mathbf{R}%
_{2},\mathbf{R}_{1};E_{F})$ and $\hat{G}(\mathbf{R}_{1},\mathbf{R}_{2};E_{F})$
mimic an outgoing spherical wave and decays as $1/R^{(d-1)/2}$ to conserve the
total probability current, so the RKKY interaction exhibits \textquotedblleft
universal\textquotedblright\ $1/R^{d}$ decay \cite{Zhang2DMater2017}, as found
in many previous studies. When the carrier's density of states on the Fermi
level vanishes, the RKKY interaction may decay even faster, e.g., when the
Fermi level locates at the Dirac/Weyl point, it follows $1/R^{3}$ decay in
graphene \cite{VozmedianoPRB2005,BreyPRL2007,SaremiPRB2007} and on the surface
of TIs \cite{LiuPRL2009,BiswasPRB2010,ZhuPRL2011,AbaninPRL2011}, and follows
$1/R^{5}$ decay in Dirac/Weyl semimetals \cite{ChangPRB2015,HosseiniPRB2015}.

\begin{figure}[tbp]
\includegraphics[width=\columnwidth]{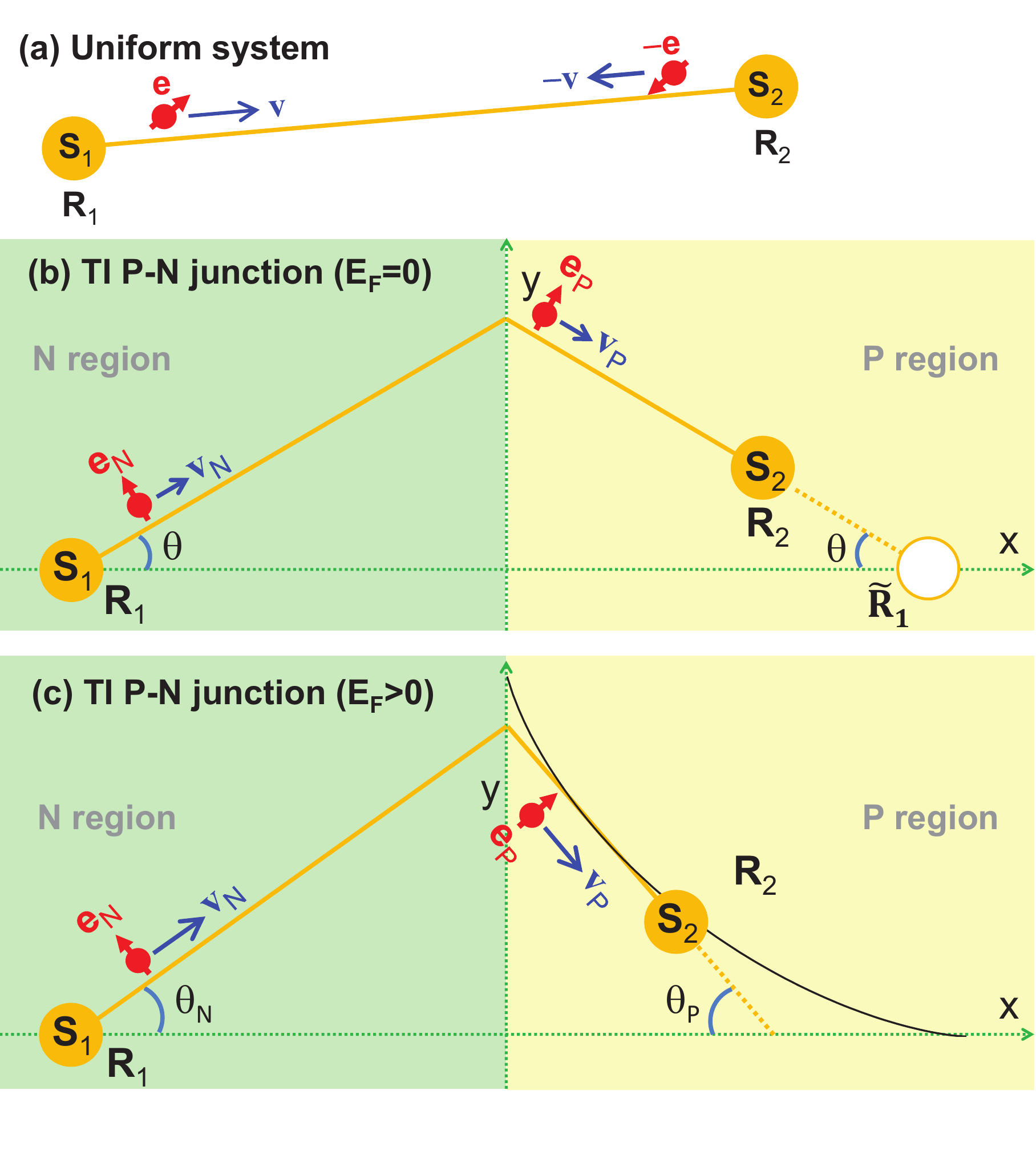} \caption{Group velocity (blue
arrows) and spin orientation (red arrows) of itinerant carriers traveling
along a classical trajectory (orange line) that connects the two local spins
(a) in uniform system, or on the surface of (b) symmetric or (c) asymmetric TI
P-N junction. The black curve in (c) indicates the caustics above the $x$
axis.}%
\label{G_SETUP}%
\end{figure}

Next we turn to the anisotropy of the RKKY\ interaction. Without spin-orbit
coupling, the carrier propagator is spin-independent, so $\hat{K}(E)$ only
contains the isotropic Heisenberg term $\hat{\mathbf{S}}_{1}\cdot
\hat{\mathbf{S}}_{2}$. With spin-orbit coupling, the anisotropy originates
intuitively as follows. In a uniform system, carriers going from
$\hat{\mathbf{S}}_{1}$ to $\hat{\mathbf{S}}_{2}$ have a group velocity along
$\mathbf{R}_{2}-\mathbf{R}_{1}$, while those going from $\hat{\mathbf{S}}_{2}$
back to $\hat{\mathbf{S}}_{1}$ have an opposite group velocity [blue arrows in
Fig. \ref{G_SETUP}(a)]. Usually, opposite group velocities $\pm\mathbf{v}$
amounts to opposite momenta $\pm\mathbf{k}$, which in turn are locked to
opposite spin orientations $\pm\mathbf{b}(\mathbf{k})$ via a generic
spin-orbit coupling $\hat{\boldsymbol{\sigma}}\cdot\mathbf{b}(\pm
\mathbf{k})=\pm\hat{\boldsymbol{\sigma}}\cdot\mathbf{b}(\mathbf{k})$ that
preserves time-reversal symmetry. For clarity, we use a unit vector
$\mathbf{e}$ ($-\mathbf{e}$) [red arrows in Fig. \ref{G_SETUP}(a)] for the
spin orientation of those carriers traveling from $\hat{\mathbf{S}}_{1}$ to
$\hat{\mathbf{S}}_{2}$ (from $\hat{\mathbf{S}}_{2}$ back to $\hat{\mathbf{S}%
}_{1}$), and $|\mathbf{n}\rangle$ for the spin-up state along the vector
$\mathbf{n}$, then $\hat{G}(\mathbf{R}_{2},\mathbf{R}_{1};E_{F})\propto
|\mathbf{e}\rangle\langle\mathbf{e}|$ and $\hat{G}(\mathbf{R}_{1}%
,\mathbf{R}_{2};E_{F})\propto|-\mathbf{e}\rangle\langle-\mathbf{e}|$, so Eq.
(\ref{IE_DEF}) gives
\[
\hat{K}(E_{F})\propto(\langle\mathbf{e}|\hat{\boldsymbol{\sigma}}%
|-\mathbf{e}\rangle\cdot\hat{\mathbf{S}}_{1})\times(\langle-\mathbf{e}%
|\hat{\boldsymbol{\sigma}}|\mathbf{e}\rangle\cdot\hat{\mathbf{S}}_{2})=\hat
{S}_{1,\mathbf{e}}^{+}\hat{S}_{2,\mathbf{e}}^{-},
\]
where $\hat{S}_{\mathbf{e}}^{+}$ ($\hat{S}_{\mathbf{e}}^{-})$ is the spin
raising (lowering) operator that increases (decreases) $\hat{\mathbf{S}}%
\cdot\mathbf{e}$ by one. Physically, this\ interaction comes from the
following two-step process [cf. Fig. \ref{G_SETUP}(a)]: \leftmargini=3mm

\begin{enumerate}
\item[(i)] A carrier with group velocity $\mathbf{v}$ and spin $|\mathbf{e}%
\rangle$ travels from $\hat{\mathbf{S}}_{1}$ to $\hat{\mathbf{S}}_{2}$ and
interacts with $\hat{\mathbf{S}}_{2}$. This interaction reverses the carrier
velocity to $-\mathbf{v}$ and the carrier spin to $|-\mathbf{e}\rangle$. It
also increases $\hat{\mathbf{S}}_{2}\cdot\mathbf{e}$ by one to conserve the
total spin along $\mathbf{e}$.

\item[(ii)] This carrier travels from $\hat{\mathbf{S}}_{2}$ back to
$\hat{\mathbf{S}}_{1}$ and interacts with $\hat{\mathbf{S}}_{1}$. This
interaction restores the carrier group velocity back to $\mathbf{v}$ and the
carrier spin back to $|\mathbf{e}\rangle$. It also decreases $\hat{\mathbf{S}%
}_{1}\cdot\mathbf{e}$ by one to conserve the total spin along $\mathbf{e}$.
\end{enumerate}

\noindent The entire process leaves the carrier spin intact, but increases
$\hat{\mathbf{S}}_{2}\cdot\mathbf{e}$ by one and decreases $\hat{\mathbf{S}%
}_{1}\cdot\mathbf{e}$ by one. Explicitly,
\begin{equation}
\hat{S}_{1,\mathbf{e}}^{+}\hat{S}_{2,\mathbf{e}}^{-}=\hat{\mathbf{S}}_{1}%
\cdot\hat{\mathbf{S}}_{2}-(\mathbf{e}\cdot\hat{\mathbf{S}}_{1})(\mathbf{e}%
\cdot\hat{\mathbf{S}}_{2})+i\mathbf{e}\cdot(\hat{\mathbf{S}}_{1}\times
\hat{\mathbf{S}}_{2}) \label{S1PS2N}%
\end{equation}
contains the isotropic Heisenberg term and the anisotropic Ising and
Dzyaloshinskii-Moriya terms, as found previously in many materials, such as
Dirac and Weyl semimetals \cite{ChangPRB2015,HosseiniPRB2015} and the surface
of TIs
\cite{LiuPRL2009,BiswasPRB2010,ZhuPRL2011,AbaninPRL2011,EfimkinPRB2014,LitvinovPRB2014,ZyuzinPRB2014}%
.

Therefore, the outgoing propagation (spin polarization) of the carriers
traveling between $\hat{\mathbf{S}}_{1}$ and $\hat{\mathbf{S}}_{2}$ generates
the $1/R^{d}$ decay (anisotropic terms) of the RKKY interaction. Usually, the
spin polarization axis $\mathbf{e}$ is determined by $\mathbf{R}%
_{2}-\mathbf{R}_{1}$ and the spin-orbit coupling [e.g., $\mathbf{e\propto
e}_{z}\times(\mathbf{R}_{2}-\mathbf{R}_{1})$ on a TI surface], making
selective control of different interactions difficult. Interestingly, the P-N
interface can change both behaviors, so it can generate and selectively
amplify anisotropic RKKY\ interactions. This physical mechanism is applicable
to any material with spin-orbit coupling. Here for clarity, we illustrate this
mechanism for the surface of a TI P-N junction.
\begin{figure*}[tbp]
\includegraphics[width=1.8\columnwidth]{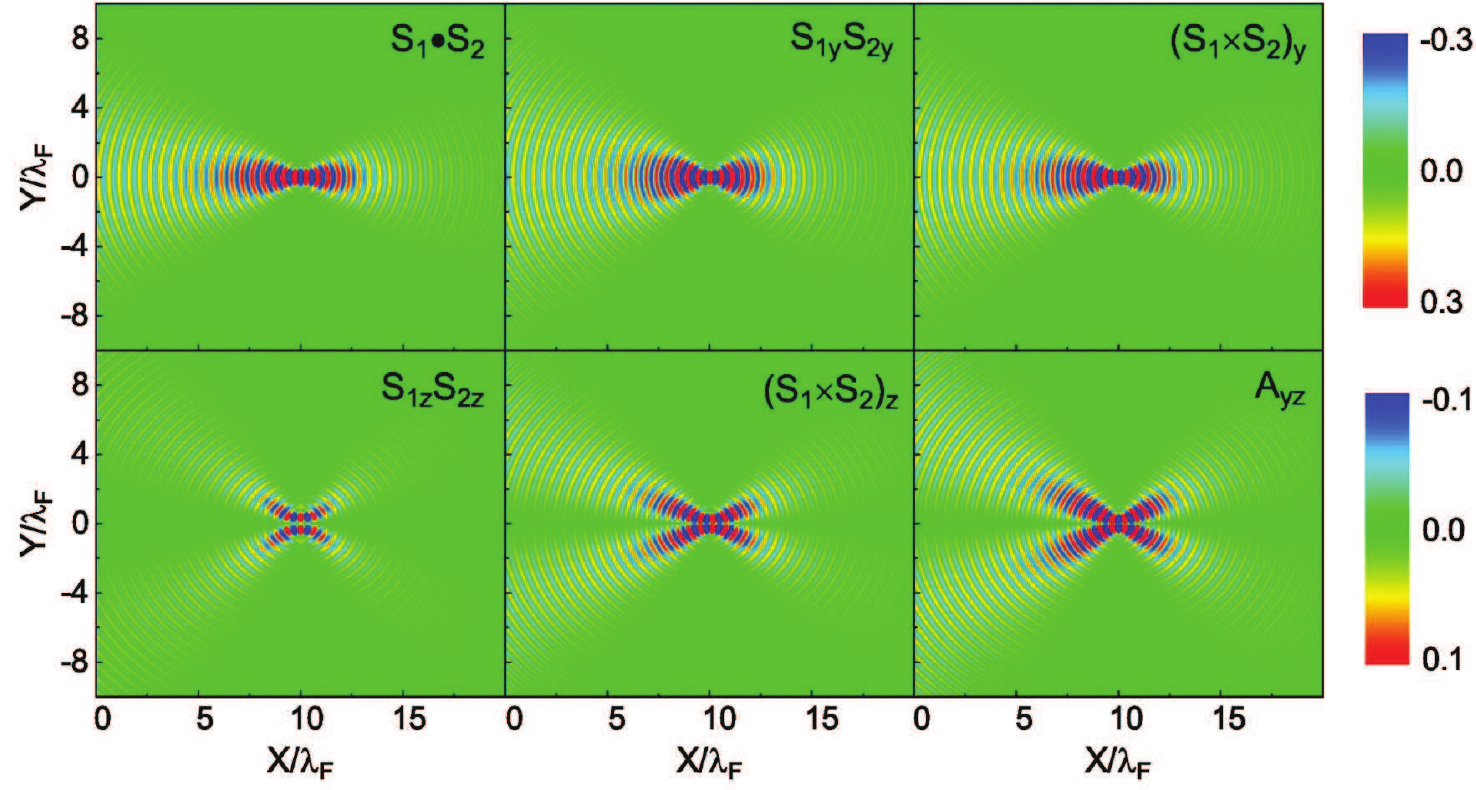}
\caption{Spatial map of different RKKY interactions [unit: $J_{0}/(2a)$] vs.
$\mathbf{R}_{2}=(X,Y)$ in a symmetric P-N junction, where $A_{\alpha\beta
}\equiv\hat{S}_{1}^{\alpha}\hat{S}_{2}^{\beta}+\hat{S}_{1}^{\beta}\hat{S}%
_{2}^{\alpha}$, the first spin is fixed at $\mathbf{R}_{1}=(-a,0)=(-10\lambda
_{F},0)$, and $J_{0}\equiv\lambda^{2}q_{F}^{2}/(8\pi^{3}v_{F})$.}
\label{G_CONTOUR_SPNJ}%
\end{figure*}

\section{RKKY interaction in TI P-N junctions}

The TI is a new class of quantum matter with extraordinary surface carriers
(in the $xy$ plane) as described by \cite{HasanRMP2010,QiRMP2011}
\[
\hat{H}_{0}=v_{F}\hat{\boldsymbol{\sigma}}\cdot(\mathbf{e}_{z}\times
\hat{\mathbf{p}}),
\]
which gives gapless linear dispersion $E_{\pm}(\mathbf{q})=\pm v_{F}%
|\mathbf{q|}$ with group velocity $\mathbf{v}_{\pm}=\pm v_{F}\mathbf{q}%
/|\mathbf{q|}$.
The corresponding eigenstate is $e^{i\mathbf{q}\cdot\mathbf{r}}|\mathbf{e}%
_{z}\times\mathbf{v}_{\pm}\rangle$, so the spin orientation of a carrier
moving with group velocity $\mathbf{v}$ is locked to $\mathbf{e}_{z}%
\times\mathbf{v}$. Very recently, there were remarkable progresses in fabricating TI-based nanostructures \cite{TongNatPhys2016, LiACSNano2015,BathonAdvMater2016,TuNatComm2016,KimACSNano2017}. In particular, the TI P-N junctions were recently fabricated
\cite{LiACSNano2015,BathonAdvMater2016,TuNatComm2016,KimACSNano2017} with
atomically abrupt interfaces \cite{KimACSNano2017}. Previous theoretical works
focus on their interesting electronic properties, such as gapless junction
states \cite{WangPRB2012}, birefringent spin lens \cite{ZhaoPRL2013}, spin
filtering \cite{HabibPRL2015}, and spin-based Mach-Zehnder interferometry
\cite{IlanPRL2015}. Here we explore a very different application: the
selective amplification of different RKKY interactions. This may be relevant
to the application of TIs to spintronics and quantum computing.

The surface carriers of a TI P-N junction is described by%
\[
\hat{H}=\hat{H}_{0}+\mathrm{sgn}(x)V_{0}%
\]
with Fermi energy $E_{F}\in\lbrack-V_{0},V_{0}]$. The left region $(x<0)$ is
N-type with electron Fermi momentum $q_{N}\equiv(V_{0}+E_{F})/v_{F}$, while
the right region $(x>0)$ is P-type with hole Fermi momentum $q_{P}\equiv
(V_{0}-E_{F})/v_{F}$. Without losing generality, we assume $\hat{\mathbf{S}%
}_{1}$ locates at $\mathbf{R}_{1}=(-a,0)$ in the N region and $\hat
{\mathbf{S}}_{2}$ locates at $\mathbf{R}_{2}=(X,Y)$ in the P region.

\subsection{Symmetric P-N junction ($E_{F}=0$)}
Here the N region and the P region have the same Fermi momenta $q_{N}%
=q_{P}=q_{F}\equiv V_{0}/v_{F}$ and Fermi wavelength $\lambda_{F}\equiv
2\pi/q_{F}$. As shown in Fig. \ref{G_SETUP}(b), an electron incident from the
N region is refracted by the P-N interface. The incident and refractive
trajectories are mirror symmetric about the P-N interface, so electrons
emanating from $\hat{\mathbf{S}}_{1}$ will be refocused by the P-N interface
onto the focal point\textit{ }$\mathbf{\tilde{R}}_{1}\equiv(a,0)$ [empty
circle in Fig. \ref{G_SETUP}(b)]. This is the well-known negative refraction of electrons across a P-N interface
\cite{CheianovScience2007,LeeNatPhys2015,ChenScience2016}, the electronic analog to its optical counterpart \cite{VeselagoSPU1968}.

On uniform TI\ surface, the RKKY interaction has rotational symmetry around
the axis $\mathbf{e\propto e}_{z}\times\mathbf{(R}_{2}-\mathbf{R}_{1})$, which
only allows three terms [Eq. (\ref{S1PS2N})] that conserve the total spin
along $\mathbf{e}$. On the surface of TI\ P-N junction, this rotational
symmetry is broken, so all interactions are allowed:\ the Heisenberg term
$\hat{\mathbf{S}}_{1}\cdot\hat{\mathbf{S}}_{2}$, the Ising terms ($\hat
{S}_{1y}\hat{S}_{2y}$ and $\hat{S}_{1z}\hat{S}_{2z}$), the
Dzyaloshinskii-Moriya terms $\hat{D}_{x},\hat{D}_{y},\hat{D}_{z}$, and
$\hat{A}_{xy},\hat{A}_{xz},\hat{A}_{yz},$ where $\hat{\mathbf{D}}\equiv
\hat{\mathbf{S}}_{1}\times\hat{\mathbf{S}}_{2}$ and $\hat{A}_{\alpha\beta}$ is
the element of the symmetric tensor operator $\mathbb{A}\equiv\hat{\mathbf{S}%
}_{1}\hat{\mathbf{S}}_{2}+\hat{\mathbf{S}}_{2}\hat{\mathbf{S}}_{1}$.
Surprisingly, numerical simulations show that only six terms (Fig.
\ref{G_CONTOUR_SPNJ}) are appreciable, while other terms are smaller by two
orders of magnitudes (not shown), suggesting the existence of a
\textit{hidden} symmetry. Figure \ref{G_CONTOUR_SPNJ} further shows that all
the terms are strongly enhanced when $\hat{\mathbf{S}}_{2}$ locates near the
focal point and different terms exhibit distinct angular dependencies. Next, we
present analytical expressions and a physical picture for these features.

For $\hat{\mathbf{S}}_{2}$ far from the focal point ($|\mathbf{R}%
_{2}-\mathbf{\tilde{R}}_{1}|\gtrsim\lambda_{F}$), the propagator from
$\hat{\mathbf{S}}_{1}$ to $\hat{\mathbf{S}}_{2}$ is dominated by the classical
trajectory [solid orange line in Fig. \ref{G_SETUP}(b)], as characterized by
an incident angle $\theta\equiv\tan^{-1}[Y/(a-X)]$, transmission amplitude
$t=\cos\theta$, and trajectory length $R_{N}=a/\cos\theta$ ($R_{P}%
=X/\cos\theta$) in the N (P) region. In the N (P) region, the carrier group
velocity is parallel (anti-parallel)\ to the momentum, so the propagation
phase is $q_{F}R_{N}$ ($-q_{F}R_{P}$). The total propagation phase is
\[
\Phi\equiv q_{F}(R_{N}-R_{P})=\mathrm{sgn}(a-X)q_{F}|\mathbf{R}_{2}%
-\mathbf{\tilde{R}}_{1}|.
\]
Using stationary phase approximation
\cite{RothPR1966,LounisPRB2011,PowerPRB2011,LiuPRB2012a} gives (see Appendix
A)
\begin{equation}
\hat{G}(\mathbf{R}_{2},\mathbf{R}_{1},E_{F})\approx t\frac{q_{F}}{iv_{F}}%
\frac{e^{i\Phi}}{\sqrt{2\pi i\Phi}}|\mathbf{e}_{\mathrm{P}}\rangle
\langle{\mathbf{e}}_{\mathrm{N}}{|},\label{G21_SPA}%
\end{equation}
where the spin orientations ${\mathbf{e}}_{\mathrm{N}},\mathbf{e}_{\mathrm{P}%
}$ are indicated by red arrows in Fig. \ref{G_SETUP}(b). The propagator from
$\hat{\mathbf{S}}_{2}$ back to $\hat{\mathbf{S}}_{1}$ is obtained from Eq.
(\ref{G21_SPA}) by time reversal of the spin (i.e., $|\mathbf{e}_{\mathrm{P}%
}\rangle\langle{\mathbf{e}}_{\mathrm{N}}{|}\rightarrow|-\mathbf{e}%
_{\mathrm{N}}\rangle\langle-{\mathbf{e}}_{\mathrm{P}}{|}$). The RKKY
interaction is dominated by the contribution on the Fermi level
\cite{PowerPRB2011,PowerPRB2012}:%
\begin{equation}
\hat{H}_{\text{\textrm{RKKY}}}\approx\frac{J^{2}v_{F}}{2\pi(R_{N}+R_{P}%
)}\operatorname{Re}\hat{K}(E_{F}).\label{RKKY_KEF}%
\end{equation}
Using the explicit expressions for the propagators gives%
\begin{equation}
\hat{H}_{\text{\textrm{RKKY}}}\approx-\frac{J_{0}t^{2}}{R_{N}+R_{P}}\frac
{2\pi}{\Phi}\operatorname{Im}(e^{2i\Phi}{\hat{S}}_{1,\mathbf{e}_{\mathrm{N}}%
}^{-}{\hat{S}}_{2,\mathbf{e}_{\mathrm{P}}}^{+})\label{RKKY_SPA}%
\end{equation}
where $J_{0}\equiv\lambda^{2}q_{F}^{2}/(8\pi^{3}v_{F})$ and ${\hat{S}%
}_{\mathbf{e}}^{+}$ (${\hat{S}}_{\mathbf{e}}^{-}$) increases
(decreases)$\ \hat{\mathbf{S}}\cdot\mathbf{e}$ by one. This interaction arises
from the following process [cf. Fig. \ref{G_SETUP}(b)]:

\leftmargini=3mm

\begin{enumerate}
\item[(1)] A carrier departs $\hat{\mathbf{S}}_{1}$ with group velocity
$\mathbf{v}_{\mathrm{N}}$ and spin $|\mathbf{e}_{\mathrm{N}}\rangle$ and
arrives at $\hat{\mathbf{S}}_{2}$ with a different group velocity
$\mathbf{v}_{\mathrm{P}}$ and spin $|\mathbf{e}_{\mathrm{P}}\rangle$, as
described by $\hat{G}(\mathbf{R}_{2},\mathbf{R}_{1},E_{F})\propto
|\mathbf{e}_{\mathrm{P}}\rangle\langle{\mathbf{e}}_{\mathrm{N}}{|}$. Next, the
contact exchange interaction with $\hat{\mathbf{S}}_{2}$ reverses the carrier
group velocity to $-\mathbf{v}_{\mathrm{P}}$ and the carrier spin to
$|-\mathbf{e}_{\mathrm{P}}\rangle$. It also increases $\hat{\mathbf{S}}%
_{2}\cdot\mathbf{e}_{\mathrm{P}}$ by one to conserve the total spin along
$\mathbf{e}_{\mathrm{P}}$.

\item[(2)] This carrier departs $\hat{\mathbf{S}}_{2}$ with group velocity
$-\mathbf{v}_{\mathrm{P}}$ and spin $|-\mathbf{e}_{\mathrm{P}}\rangle$ and
travels back to $\hat{\mathbf{S}}_{1}$ with a different group velocity
$-\mathbf{v}_{\mathrm{N}}$ and spin $|-\mathbf{e}_{\mathrm{N}}\rangle$, as
described by $\hat{G}(\mathbf{R}_{1},\mathbf{R}_{2},E_{F})\propto
|-\mathbf{e}_{\mathrm{N}}\rangle\langle-{\mathbf{e}}_{\mathrm{P}}{|}$. Next,
the contact exchange interaction with $\hat{\mathbf{S}}_{1}$ restores its
initial velocity $\mathbf{v}_{\mathrm{N}}$ and initial spin $|\mathbf{e}%
_{\mathrm{N}}\rangle$. It also decreases $\hat{\mathbf{S}}_{1}\cdot
\mathbf{e}_{\mathrm{N}}$ by one to conserve the total spin along
$\mathbf{e}_{\mathrm{N}}$.
\end{enumerate}

We use a complex vector $\mathbf{g}=(0,1,i\sin\theta)$ and $g_{0}%
\equiv(\mathbf{g}\cdot\mathbf{g})^{1/2}$ to characterize the spin
polarization: $|\mathbf{e}_{\mathrm{P}}\rangle\langle{\mathbf{e}}_{\mathrm{N}%
}{|=(g}_{0}+\hat{\boldsymbol{\sigma}}\cdot\mathbf{g})/2$, then Eq.
(\ref{RKKY_SPA}) becomes
\begin{equation}
\hat{H}_{\text{\textrm{RKKY}}}\approx-\frac{J_{0}t^{2}}{R_{N}+R_{P}}\frac
{2\pi}{\Phi}\operatorname{Im}e^{2i\Phi}\left(  g_{0}^{2}\hat{\mathbf{S}}%
_{1}\cdot\hat{\mathbf{S}}_{2}+ig_{0}\mathbf{g}\cdot\hat{\mathbf{D}}%
-\mathbf{g}\cdot\mathbb{A}\cdot\mathbf{g}\right)  , \label{RKKY_SPA1}%
\end{equation}
which agrees with our numerical results in Fig. \ref{G_CONTOUR_SPNJ} at
$|\mathbf{R}_{2}-\mathbf{\tilde{R}}_{1}|\gtrsim\lambda_{F}$, e.g., the
enhancement of the RKKY interaction near the focal point originates from
$2\pi/\Phi$, while the distinct angular dependencies of different interactions
are well described by:%
\[
\left\{
\begin{array}
[c]{cc}%
\hat{\mathbf{S}}_{1}\cdot\hat{\mathbf{S}}_{2} & \sim\cos^{5}\varphi,\\
\hat{S}_{1y}\hat{S}_{2y} & \sim\cos^{3}\varphi,\\
\hat{D}_{y} & \sim\cos^{4}\varphi,
\end{array}
\right.  \ \ \ \ \left\{
\begin{array}
[c]{cc}%
\hat{S}_{1z}\hat{S}_{2z} & \sim\cos^{3}\varphi\sin^{2}\varphi,\\
\hat{D}_{z} & \sim\cos^{4}\varphi\sin\varphi,\\
\hat{A}_{yz} & \sim\cos^{3}\varphi\sin\varphi.
\end{array}
\right.
\]
where $\varphi$ is the polar angle of $\mathbf{R}_{2}-\mathbf{\tilde{R}}_{1}$.

For comparison, on a uniform N-type TI surface with Fermi momentum $q_{F}$,
the classical trajectory going from $\hat{\mathbf{S}}_{1}$ to $\hat
{\mathbf{S}}_{2}$ is a straight line with a length $R\equiv|\mathbf{R}%
_{2}-\mathbf{R}_{1}|$ [cf. Fig. \ref{G_SETUP}(a)], so the propagator is
obtained from Eq. (\ref{G21_SPA}) by setting $t=1$, $\mathbf{e}_{\mathrm{N}%
}=\mathbf{e}_{\mathrm{P}}=\mathbf{e}\propto\mathbf{e}_{z}\times(\mathbf{R}%
_{2}-\mathbf{R}_{1})$, and $\Phi=q_{F}R$, while the RKKY\ interaction is
obtained from Eq. (\ref{RKKY_SPA}) by further replacing $R_{N}+R_{P}$ with
$R$. Compared with their counterparts on a uniform TI surface, Eqs.
(\ref{G21_SPA}) and (\ref{RKKY_SPA}) exhibit two distinguishing features.
First, the carrier propagators depend on $\mathbf{R}_{2}-\mathbf{\tilde{R}%
}_{1}$ instead of $\mathbf{R}_{2}-\mathbf{R}_{1}$, as if $\hat{\mathbf{S}}%
_{1}$ located at $\mathbf{\tilde{R}}_{1}$ instead of $\mathbf{R}_{1}$. This
spatial symmetry is \textit{absent} from the system Hamiltonian. Due to this
symmetry, when $|\mathbf{R}_{2}-\mathbf{\tilde{R}}_{1}|$ is fixed, the
RKKY\ interaction in Eq. (\ref{RKKY_SPA}) decays as $1/(R_{N}+R_{P})$, in
contrast to the much faster $1/R^{d}$ decay in uniform $d$-dimensional
systems. Second, a carrier departs $\hat{\mathbf{S}}_{1}$ with spin
orientation $\mathbf{e}_{\mathrm{N}}$ and arrives at $\hat{\mathbf{S}}_{2}$
with a different spin orientation $\mathbf{e}_{\mathrm{P}}$. This lifts the
constraint of total spin conservation along $\mathbf{e}$ on a uniform TI
surface [cf. Eq. (\ref{S1PS2N})]. Interestingly, $\mathbf{e}_{\mathrm{N}}$ and
$\mathbf{e}_{\mathrm{P}}$ are symmetric about the $x$ axis. This spin symmetry
is also \textit{absent} from the system Hamiltonian. Due to this symmetry, Eq.
(\ref{RKKY_SPA1}) only contains six terms as shown in Fig.
\ref{G_CONTOUR_SPNJ}: other terms are high-order contributions beyond the
stationary phase approximation and hence are much smaller. This
\textit{hidden} spatial symmetry and spin symmetry arise from the mirror
reflection symmetry of the electron and hole Fermi contours of a symmetric P-N
junction \cite{ZhangPRB2016}.

The results above are valid for $\hat{\mathbf{S}}_{2}$ far from
$\mathbf{\tilde{R}}_{1}$, otherwise
\begin{equation}
\hat{G}(\mathbf{R}_{2},\mathbf{R}_{1},E_{F})=\frac{q_{F}}{4\pi iv_{F}}%
(\tilde{g}_{0}+\hat{\boldsymbol{\sigma}}\cdot\mathbf{\tilde{g}}%
),\label{G_FOCAL}%
\end{equation}
where $\tilde{g}_{0}$ and $\mathbf{\tilde{g}}\equiv(0,\tilde{g}_{y},\tilde
{g}_{z})$ are oscillatory functions of $q_{F}(\mathbf{R}_{2}-\mathbf{\tilde
{R}}_{1})$ (see Appendix A). Then Eq. (\ref{RKKY_KEF}) gives%
\begin{equation}
\hat{H}_{\mathrm{RKKY}}\approx-\frac{J_{0}}{2a}\operatorname{Re}\left(
\frac{\tilde{g}_{0}^{2}+\mathbf{\tilde{g}}\cdot\mathbf{\tilde{g}}}{2}%
\hat{\mathbf{S}}_{1}\cdot\hat{\mathbf{S}}_{2}+i\tilde{g}_{0}\mathbf{\tilde{g}%
}\cdot\hat{\mathbf{D}}-\mathbf{\tilde{g}}\cdot\mathbb{A}\cdot\mathbf{\tilde
{g}}\right)  .\label{RKKY_NF}%
\end{equation}
Equations (\ref{RKKY_SPA1}) and (\ref{RKKY_NF}) are complementary and provide
a full description for the spatial map of the RKKY interaction. When
$\mathbf{R}_{2}-\mathbf{\tilde{R}}_{1}$ and hence $\tilde{g}_{0}%
,\mathbf{\tilde{g}}$ are fixed, the RKKY\ interaction follows $1/R$ decay,
instead of the much faster $1/R^{d}$ decay in $d$-dimensional uniform systems.
When $\hat{\mathbf{S}}_{2}$ locates at the focal point $\mathbf{\tilde{R}}%
_{1}$, we have $\tilde{g}_{0}=\pi/2$, $\tilde{g}_{y}=2$, and $\tilde{g}_{z}%
=0$, so
\[
\hat{H}_{\mathrm{RKKY}}\approx-\frac{J_{0}}{2a}\left(  \frac{\pi^{2}/4+4}%
{2}\hat{\mathbf{S}}_{1}\cdot\hat{\mathbf{S}}_{2}-4\hat{S}_{1}^{y}\hat{S}%
_{2}^{y}\right)  .
\]
Compared with a uniform TI surface with the same carrier concentration \cite{ZhuPRL2011}, the
P-N interface amplifies the RKKY interaction by a dramatic factor
$\sim2R/\lambda_{F}$ that increases with increasing distance $R=2a$.

\subsection{Asymmetric P-N junction ($E_{F}\neq0$)}

\begin{figure*}[ptb]
\includegraphics[width=1.8\columnwidth]{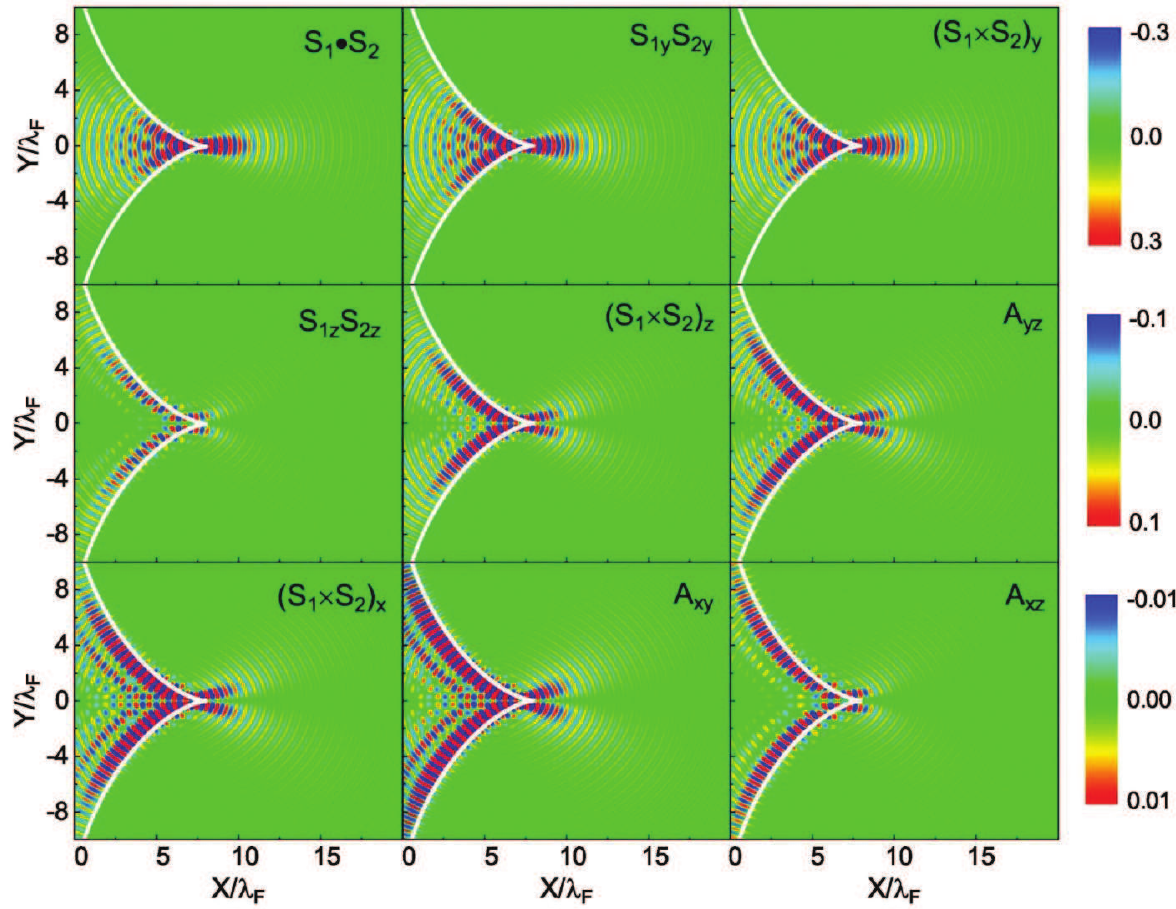}
\includegraphics[width=1.8\columnwidth]{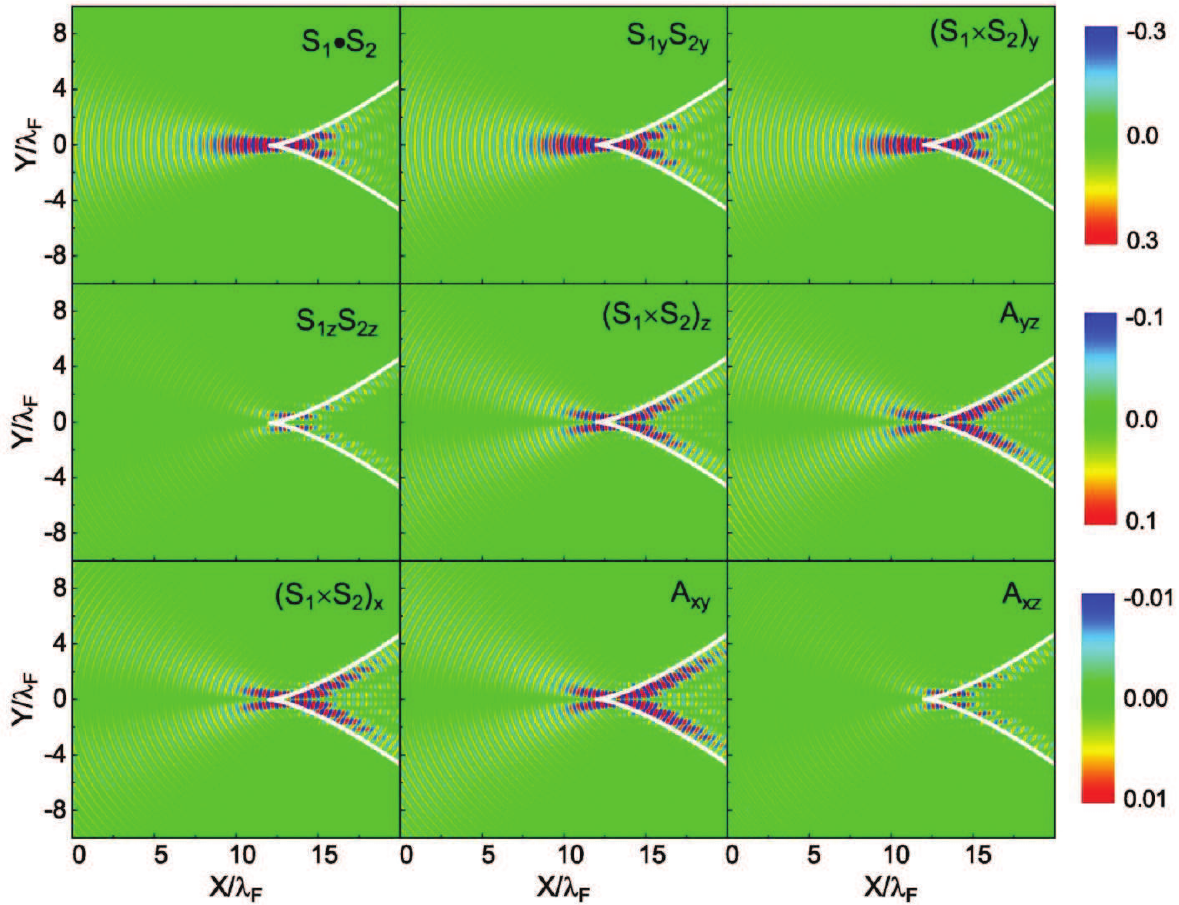}
\caption{Numerical results for the spatial map of the RKKY interactions [unit:
$J_{0}/(2a)$] as functions of $\mathbf{R}_{2}=(X,Y)$ in the P region of a P-N
junction. The first spin is fixed at $\mathbf{R}_{1}=(-a,0)=(-10\lambda_{F},0)$. The white lines indicate the
caustics. Upper part: $E_{F}=V_{0}/9$ and hence $n=0.8$. Lower part: $E_{F}=-V_{0}/11$ and hence $n=1.2$.}%
\label{G_RKKY_CONTOUR}%
\end{figure*}Here $q_{N}\neq q_{P}$, so the P-N junction is characterized by
two parameters: $q_{F}\equiv(q_{N}+q_{P})/2=V_{0}/v_{F}$ (or equivalently
$\lambda_{F}\equiv2\pi/q_{F}$) controls the average carrier density, while the
effective \textquotedblleft refractive\textquotedblright\ index $n\equiv
q_{P}/q_{N}$ controls the degree of asymmetry of the junction. As shown in
Fig. \ref{G_SETUP}(c), the incident and refractive trajectories are no longer
mirror symmetric about the P-N interface. The electrons emanating from
$\hat{\mathbf{S}}_{1}$ are refocused imperfectly by the P-N interface, giving
rise to \textit{caustics} [black curve in Fig. \ref{G_SETUP}(c)]
\cite{CheianovScience2007}, as described by
\begin{align*}
Y_{\mathrm{cau}}(X)  &  \equiv\pm\frac{\lbrack(an)^{2/3}-X^{2/3}]^{3/2}}%
{\sqrt{1-n^{2}}}\ \ (\mathrm{for}\ n<1),\\
Y_{\mathrm{cau}}(X)  &  \equiv\pm\frac{\lbrack X^{2/3}-(na)^{2/3}]^{3/2}%
}{\sqrt{n^{2}-1}}\ \ (\mathrm{for\ }n>1).
\end{align*}
For weak asymmetry (e.g., $n=0.8$ and $n=1.2$), the RKKY interactions are
shown in Fig. \ref{G_RKKY_CONTOUR}. For $n=0.8$ ($n=1.2$), all the
interactions are enhanced when $\hat{\mathbf{S}}_{2}$ locates in the left
(right) neighborhood of the caustics, but are suppressed in the right (left)
neighborhood. Moreover, the interactions $\hat{D}_{x}$, $\hat{A}_{xy}$, and
$\hat{A}_{xz}$ are significantly smaller than other terms. Next, we present
analytical expressions that reproduce these features.

As shown in Fig. \ref{G_SETUP}(c), the classical trajectory from
$\hat{\mathbf{S}}_{1}$ to $\hat{\mathbf{S}}_{2}$ is characterized by an
incident angle $\theta_{N}$, refractive angle $\theta_{P}$, transmission
amplitude $t=\cos\theta_{N}/\cos[(\theta_{P}-\theta_{N})/2]$, and trajectory
length $R_{N}\equiv a/\cos\theta_{N}$ ($R_{P}\equiv X/\cos\theta_{P}$) in the
N (P) region, so that the total propagation phase is $\Phi=q_{N}R_{N}%
-q_{P}R_{P}$. Here $\theta_{N}$ and $\theta_{P}$ are determined by the
momentum conservation along the $y$ axis, $\sin\theta_{N}=n\sin\theta_{P}$,
and the geometric constraint $Y=a\tan\theta_{N}-X\tan\theta_{P}$. For
$\hat{\mathbf{S}}_{2}$ far from the caustics, the propagator is dominated by a
single classical trajectory (see Appendix B):
\begin{equation}
\hat{G}(\mathbf{R}_{2},\mathbf{R}_{1},E_{F})\approx t\frac{q_{F}}{iv_{F}}%
\frac{e^{i\Phi}}{\sqrt{2\pi iq_{F}\tilde{R}}}|\mathbf{e}_{\mathrm{P}}%
\rangle\langle\mathbf{e}_{\mathrm{N}}|,\label{G_ASYM}%
\end{equation}
where $\mathbf{e}_{\mathrm{N}},\mathbf{e}_{\mathrm{P}}$ are shown in Fig.
\ref{G_SETUP}(c), and
\begin{equation}
\tilde{R}\equiv\frac{1+n}{2}R_{N}-\frac{1+1/n}{2}\frac{\cos^{2}\theta_{N}%
}{\cos^{2}\theta_{P}}R_{P}.\label{RT}%
\end{equation}
Taking the time reversal of the carrier spin (i.e., $|\mathbf{e}_{\mathrm{P}%
}\rangle\langle\mathbf{e}_{\mathrm{N}}|\rightarrow|-\mathbf{e}_{\mathrm{N}%
}\rangle\langle-\mathbf{e}_{\mathrm{P}}|$) gives the propagator from
$\hat{\mathbf{S}}_{2}$ back to $\hat{\mathbf{S}}_{1}$. Substituting the
propagators into Eq. (\ref{RKKY_KEF}) gives%
\[
\hat{H}_{\mathrm{RKKY}}\approx-\frac{t^{2}J_{0}}{R_{N}+R_{P}}\frac{2\pi}%
{q_{F}\tilde{R}}\operatorname{Im}e^{2i\Phi}{\hat{S}}_{1,\mathbf{e}%
_{\mathrm{N}}}^{-}{\hat{S}}_{2,\mathbf{e}_{\mathrm{P}}}^{+}.
\]
The physical process leading to this interaction is similar to that for a
symmetric P-N junction, except that $\mathbf{e}_{\mathrm{N}}$ and
$\mathbf{e}_{\mathrm{P}}$ are no longer symmetric about the $x$ axis.

\begin{figure}[ptb]
\includegraphics[width=\columnwidth]{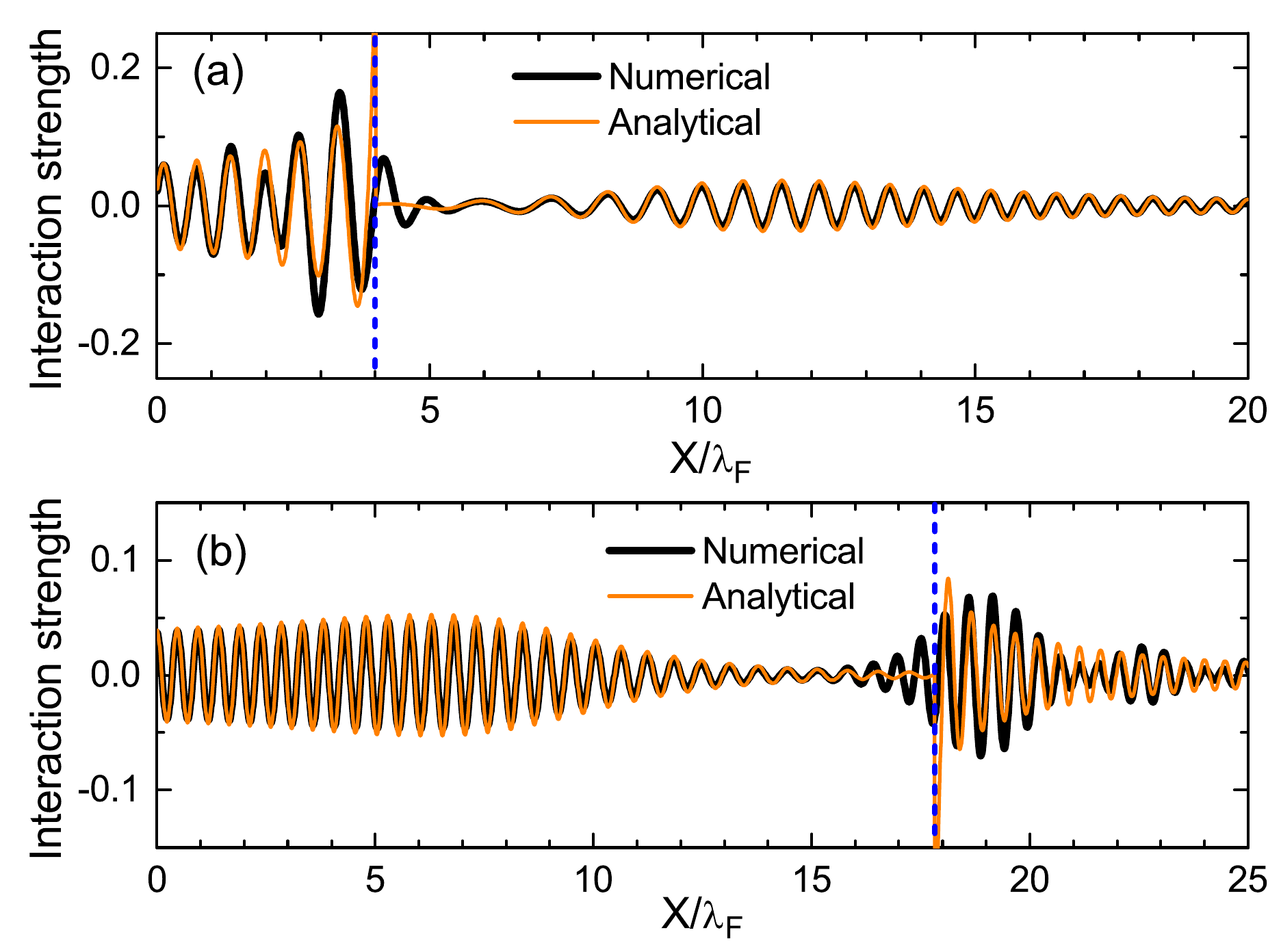}\caption{Strength of the
interaction $\hat{A}_{yz}$ [unit: $J_{0}/(2a)$] as a function of $X$ (with
$Y=3\lambda_{F}$ fixed) in an asymmetric TI P-N junction with\ (a) $n=0.8$ and
(b) $n=1.2$. The vertical gray lines marks the caustics.}%
\label{G_COMPARE}%
\end{figure}

We use a complex vector $\mathbf{g}=(\sin[(\theta_{P}-\theta_{N}%
)/2],\cos[(\theta_{P}-\theta_{N})/2],i\sin[\theta_{N}+\theta_{P})/2])$ and
$g_{0}=(\mathbf{g}\cdot\mathbf{g})^{1/2}$ to characterize the spin
polarization: $|\mathbf{e}_{\mathrm{P}}\rangle\langle{\mathbf{e}}_{\mathrm{N}%
}{|=(g}_{0}+\hat{\boldsymbol{\sigma}}\cdot\mathbf{g})/2$, then%
\begin{equation}
\hat{H}_{\text{\textrm{RKKY}}}\approx-\frac{J_{0}t^{2}}{R_{N}+R_{P}}\frac
{2\pi}{q_{F}\tilde{R}}\operatorname{Im}e^{2i\Phi}(g_{0}^{2}\hat{\mathbf{S}%
}_{1}\cdot\hat{\mathbf{S}}_{2}+ig_{0}\mathbf{g}\cdot\hat{\mathbf{D}%
}-\mathbf{g}\cdot\mathbb{A}\cdot\mathbf{g}), \label{RKKY_SPA_10TERM}%
\end{equation}
which agrees with our numerical results in Fig. \ref{G_RKKY_CONTOUR} as long
as $\hat{\mathbf{S}}_{2}$ is away from the caustics by a distance
$\gtrsim\lambda_{F}$, see Fig. \ref{G_COMPARE} for a comparison. For weak
asymmetry, i.e., $|E_{F}|\ll V_{0}$ and hence $n\approx1$, the difference
$\theta_{N}-\theta_{P}$ is small, so the polarization vector $\mathbf{g}$ has
a small $x$ component. This explains why $D_{x}$ and $A_{xy},A_{xz}$ in Fig.
\ref{G_RKKY_CONTOUR} are significantly smaller than other terms.

\subsection{Selective generation and amplification of RKKY\ interactions}

\begin{figure*}[ptb]
\includegraphics[width=1.8\columnwidth]{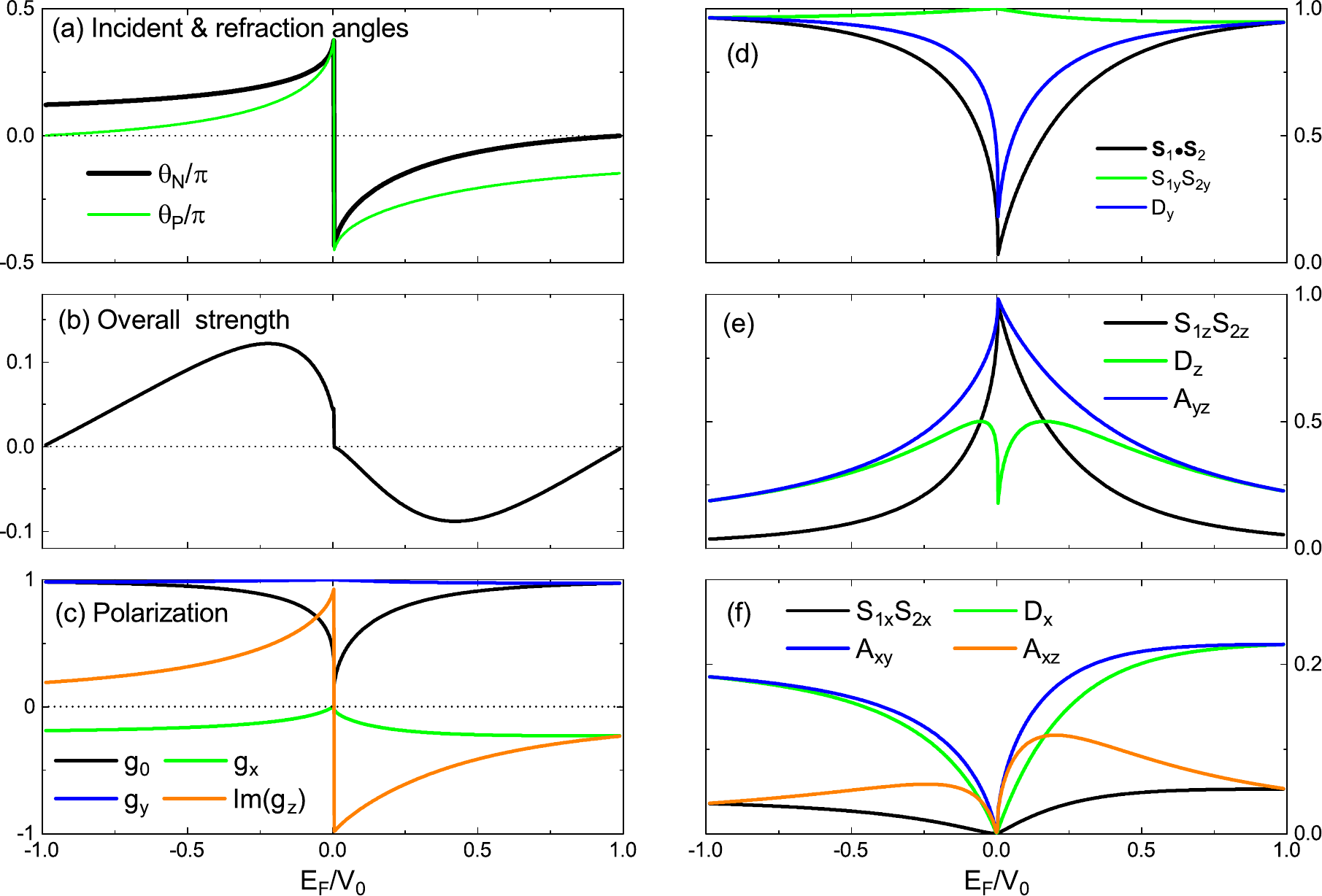}
\caption{(a) Incident and refraction angles, (b) overall strength
$(2\pi/\tilde{\Phi})J_{0}t^{2}/(R_{N}+R_{P})$ [unit: $J_{0}/(2a)]$, (c) spin
polarization vector, and (d)-(f) envelopes of different interaction terms
[unit: $J_{0}/(2a)]$ as functions of $E_{F}/V_{0}$. The two local spins are
fixed at $\mathbf{R}_{1}=(-5\lambda_{F},0)$ and $\mathbf{R}_{2}=(4\lambda
_{F},2\lambda_{F})$. The dashed lines in (a)-(c) are guides to the eye.}%
\label{G_RKKYVSEF}%
\end{figure*}

The TI\ P-N junction is controlled by two parameters: (i) The junction voltage
$V_{0}$ or equivalently the average Fermi momentum $q_{F}\equiv V_{0}/v_{F}$
(or $\lambda_{F}\equiv2\pi/q_{F}$) characterizes the average carrier density.
(ii) The Fermi level $E_{F}$ or equivalently the \textquotedblleft
refractive\textquotedblright\ index $n$ characterizes the degree of asymmetry
of the\ junction. In Eq. (\ref{RKKY_SPA_10TERM}), except for $J_{0}\propto
q_{F}^{2}$ and $\Phi\propto q_{F}$, other quantities ($\theta_{N},\theta
_{P},t,\tilde{R},\mathbf{g},$ and $g_{0}$) are completely determined by the
locations of $\mathbf{S}_{1}$ and $\mathbf{S}_{2}$ and the \textquotedblleft
refraction\textquotedblright\ index $n$. Therefore, when the locations of
$\mathbf{S}_{1}$ and $\mathbf{S}_{2}$ are fixed, $q_{F}$ controls the overall
strength $(2\pi/\tilde{\Phi})J_{0}t^{2}/(R_{N}+R_{P})\propto q_{F}$ of all the
interactions, while $n$ controls the spin polarization vector $\mathbf{g}$ and
hence the generation and selective amplification of different interactions.

Specifically, setting $n=1$ gives a symmetric P-N junction with mirror
symmetry on the Fermi contour, so $\theta_{N}=\theta_{P}=\theta$,
$\mathbf{g}=(0,1,i\sin\theta)$, and the RKKY interaction is dominated by six
terms [Eq. (\ref{RKKY_SPA1})]. Tuning $n$ slightly away from unity gives rise
to small $\theta_{N}-\theta_{P}$ and hence $g_{x}$, which in turn generates
three weak interactions $\hat{D}_{x}$, $\hat{A}_{xz}$, and $\hat{A}_{xy}$.
Further tuning $n$ far from unity increases the degree of asymmetry of the P-N
junction and hence further enhances the strengths of these interactions. As an
example, we consider $a=5\lambda_{F}$, $X=4\lambda_{F}$, and $Y=2\lambda_{F}$,
and vary $E_{F}/V_{0}$ from $-1$ across zero to $+1$ to tune the
\textquotedblleft refraction\textquotedblright\ index $n$ from $+\infty$
across $+1$ to $0$. This variation changes the incident and refraction angles
[Fig. \ref{G_RKKYVSEF}(a)] and hence leads to two effects. First, it changes
the effective distances $\tilde{R}$ and $R_{N}+R_{P}$ [Fig. \ref{G_RKKYVSEF}%
(b)] and hence changes the overall strength of all the interactions, as shown
in Fig. \ref{G_RKKYVSEF}(c). Second, it changes the spin polarization vector
$\mathbf{g}$ [Fig. \ref{G_RKKYVSEF}(d)] and hence tunes the anisotropy of the
RKKY\ interactions. According to Eq. (\ref{RKKY_SPA_10TERM}), each interaction
term oscillates rapidly as $\sin(2\Phi)$ or $\cos(2\Phi)$ with a corresponding
envelope, so it is better to quantify the anisotropy of different interactions
by these envelopes. Then Fig. \ref{G_RKKYVSEF}(d)-(f) clearly demonstrates the
possibility to tune the relative strength of different interactions by tuning
the \textquotedblleft refraction\textquotedblright\ index.

\section{Conclusions}

We have proposed a physical mechanism for the selective generation and
amplification of anisotropic RKKY\ interactions. The key is to utilize the
negative refraction across a P-N junction to amplify all the interactions, and
utilize the deflection of the carrier spin by the P-N interface (via the
spin-orbit coupling) to achieve selected generation and amplification of
anisotropic terms. Specifically, the junction potential $V_{0}$ controls the
average carrier density and hence the overall strength of all the
interactions, while the Fermi energy controls the degree of asymmetry of the
P-N\ junction and hence the generation and amplification of anisotropic
interactions. Although we have limited our numerical and analytical
discussions to the surface of a TI\ P-N junction for the sake of specificity,
this physical mechanism is applicable to the P-N junction of an arbitrary
system with spin-orbit coupling. Compared with previous works that focus on
the dependence of the RKKY\ interaction on specific materials and carrier
density, our work identifies the P-N interface as a second knob which, when
combined with the carrier density, may open up the possibility for the
independent control of the strength and anisotropy of the RKKY interaction in
spin-orbit coupled systems.

\begin{acknowledgments}
This work was supported by the National Key R\&D Program of China (Grants No.
2017YFA0303400), the MOST of China (Grants No. 2014CB848700), the NSFC (Grants
No. 11774021, No. 11504018, and No. 11404043), and the NSFC program for
\textquotedblleft Scientific Research Center\textquotedblright\ (Grant No.
U1530401). We acknowledge the computational support from the Beijing
Computational Science Research Center (CSRC).
\end{acknowledgments}

\appendix{}

\section{Carrier propagators in symmetric TI\ P-N junction}

Given the carrier Hamiltonian $\hat{H}$, the propagator of the carriers is
defined as $G_{\mu,\nu}(\mathbf{r},\mathbf{r}_{0},E_{F})\equiv\langle
\mathbf{r},\mu|(E_{F}-\hat{H}+i0^{+})^{-1}|\mathbf{r}_{0},\nu\rangle$, where
$\mu,\nu=+$ (spin up) or $-$ (spin down) denotes the spin states. When
$\hat{H}$ is invariant under time-reversal operation $\hat{\theta}$, i.e.,
$\hat{\theta}\hat{H}\hat{\theta}^{-1}=\hat{H}$, we can use $\langle i|\hat
{O}|j\rangle=\langle\hat{\theta}i|\hat{\theta}\hat{O}\hat{\theta}^{-1}%
|\hat{\theta}j\rangle^{\ast}=\langle\hat{\theta}j|(\hat{\theta}\hat{O}%
\hat{\theta}^{-1})^{\dagger}|\hat{\theta}i\rangle$ to obtain%
\[
G_{\mu\nu}(\mathbf{r},\mathbf{r}_{0},E_{F})=\langle\hat{\theta}(\mathbf{r}%
_{0},\nu)|(E_{F}-\hat{H}+i0^{+})^{-1}|\hat{\theta}(\mathbf{r},\mu)\rangle,
\]
where $|\hat{\theta}(\mathbf{r},\pm)\rangle$ is the time-reversal of
$|\mathbf{r},\pm\rangle$. Using $\hat{\theta}|\mathbf{r}\rangle=\hat{\theta
}|\mathbf{r}\rangle$ and $\hat{\theta}|\pm\rangle=\pm|\mp\rangle$, we obtain
$|\hat{\theta}(\mathbf{r},\pm)\rangle=\pm|\mathbf{r},\mp\rangle$ and hence%
\begin{align*}
G_{\uparrow\uparrow}(\mathbf{r},\mathbf{r}_{0},E_{F})  &  =G_{\downarrow
\downarrow}(\mathbf{r}_{0},\mathbf{r},E_{F}),\\
G_{\downarrow\downarrow}(\mathbf{r},\mathbf{r}_{0},E_{F})  &  =G_{\uparrow
\uparrow}(\mathbf{r}_{0},\mathbf{r},E_{F}),\\
G_{\uparrow\downarrow}(\mathbf{r},\mathbf{r}_{0},E_{F})  &  =-G_{\downarrow
\uparrow}(\mathbf{r}_{0},\mathbf{r},E_{F}),\\
G_{\downarrow\uparrow}(\mathbf{r},\mathbf{r}_{0},E_{F})  &  =-G_{\uparrow
\downarrow}(\mathbf{r}_{0},\mathbf{r},E_{F}).
\end{align*}
Namely, given the $2\times2$ propagator $\hat{G}(\mathbf{r},\mathbf{r}%
_{0},E_{F})=G_{0}+\mathbf{G}\cdot\hat{\boldsymbol{\sigma}}$ from
$\mathbf{r}_{0}$ to $\mathbf{r}$, the $2\times2$ propagator from $\mathbf{r}$
back to $\mathbf{r}_{0}$ can be obtained by taking time reversal of the
carrier spin: $\hat{G}(\mathbf{r}_{0},\mathbf{r},E_{F})=G_{0}-\mathbf{G}%
\cdot\hat{\boldsymbol{\sigma}}$. Therefore, in the following, we only consider
the propagator from $\mathbf{R}_{1}$ to $\mathbf{R}_{2}$.

For convenience, we use $\mathbf{e}_{\varphi}\equiv\cos\varphi\mathbf{e}%
_{x}+\sin\varphi\mathbf{e}_{y}$ to denote a unit vector with polar angle
$\varphi$. For $E_{F}=0$, the N region and the P region has the same Fermi
momentum $q_{F}\equiv V_{0}/v_{F}$. We consider a right-going incident
electron from the N region with an incident angle $\varphi$, momentum
$(q_{x},q_{y})\equiv q_{F}\mathbf{e}_{\varphi}$, group velocity $\mathbf{v}%
_{\mathrm{N}}\equiv v_{F}\mathbf{e}_{\varphi}$, and spin state
\[
|\mathbf{e}_{z}\times\mathbf{v}_{\mathrm{N}}\rangle=\frac{1}{\sqrt{2}}%
\begin{bmatrix}
e^{-i\varphi/2}\\
ie^{i\varphi/2}%
\end{bmatrix}
,
\]
where $q_{x}\equiv(q_{F}^{2}-q_{y}^{2})^{1/2}$. The reflection electron has a
momentum $(-q_{x},q_{y})\equiv q_{F}\mathbf{e}_{\pi-\varphi}$, group velocity
$\mathbf{v}_{\mathrm{r}}=v_{F}\mathbf{e}_{\pi-\varphi}$, and spin state
\[
|\mathbf{e}_{z}\times\mathbf{v}_{\mathrm{r}}\rangle=\frac{1}{\sqrt{2}}%
\begin{bmatrix}
e^{i\varphi/2}\\
-ie^{-i\varphi/2}%
\end{bmatrix}
.
\]
The transmission electron has a momentum $(-q_{x},q_{y})=q_{F}\mathbf{e}%
_{\pi-\varphi}$, group velocity $\mathbf{v}_{\mathrm{P}}\equiv v_{F}%
\mathbf{e}_{-\varphi}$, and spin state
\[
|\mathbf{e}_{z}\times\mathbf{v}_{\mathrm{P}}\rangle=\frac{1}{\sqrt{2}}%
\begin{bmatrix}
e^{i\varphi/2}\\
ie^{-i\varphi/2}%
\end{bmatrix}
.
\]
From the continuity equation at the P-N\ interface $x=0$,%
\[
|\mathbf{e}_{z}\times\mathbf{v}_{\mathrm{N}}\rangle+r|\mathbf{e}_{z}%
\times\mathbf{v}_{\mathrm{r}}\rangle=t|\mathbf{e}_{z}\times\mathbf{v}%
_{\mathrm{P}}\rangle,
\]
we obtain the transmission amplitude $t(\varphi)=\cos\varphi$ and the
reflection amplitude $r(\varphi)=i\sin\varphi$.

The carrier propagator from $\mathbf{R}_{1}=(-a,0)\ $in the N region to
$\mathbf{R}_{2}=(X,Y)$ in the P region is given by
\[
\hat{G}(\mathbf{R}_{2},\mathbf{R}_{1},E_{F})=\int_{-\infty}^{\infty}%
\frac{dq_{y}}{2\pi}t(\varphi)e^{i\phi(q_{y})}\frac{|\mathbf{e}_{z}%
\times\mathbf{v}_{\mathrm{P}}\rangle{\langle\mathbf{e}}_{z}\times
\mathbf{v}_{\mathrm{N}}{|}}{iv_{F}q_{x}/q_{F}},
\]
where $\phi(q_{y})=q_{y}Y+q_{x}(a-X)$. By keeping traveling waves only, we can
replace $\int_{-\infty}^{\infty}dq_{y}$ by $\int_{-q_{F}}^{q_{F}}dq_{y}$ and
obtain Eq. (\ref{G_FOCAL}) in the main text, where%
\begin{align*}
\tilde{g}_{0} &  \equiv\int_{-\pi/2}^{\pi/2}\cos^{2}\varphi e^{i\phi(\varphi
)}d\varphi,\\
\tilde{g}_{y} &  \equiv\int_{-\pi/2}^{\pi/2}\cos\varphi e^{i\phi(\varphi
)}d\varphi,\\
\tilde{g}_{z} &  \equiv i\int_{-\pi/2}^{\pi/2}\sin\varphi\cos\varphi
e^{i\phi(\varphi)}d\varphi,
\end{align*}
and $\phi(\varphi)=q_{F}[Y\sin\varphi+(a-X)\cos\varphi]$.

The incident angle $\theta$ of the \textit{classical} trajectory is determined
by $\partial_{\varphi}\phi(\varphi)=0$ as $\theta=\tan^{-1}[Y/(a-X)]$. In
terms of $\theta$ and the total propagation phase $\Phi\equiv q_{F}%
(a-X)/\cos\theta$ along the classical trajectory, we have $\phi(\varphi
)=\Phi\cos(\varphi-\theta)$. For $\mathbf{R}_{2}$ far from the focal point
$\mathbf{\tilde{R}}_{1}\equiv(a,0)$, we can replace $\phi(\varphi)$ by its
Taylor expansion near $\varphi=\theta$ up to the second order to obtain
\[
(\tilde{g}_{0},\tilde{g}_{y},\tilde{g}_{z})=\cos\theta\sqrt{\frac{2\pi}{i\Phi
}}e^{i\Phi}(\cos\theta,1,i\sin\theta)
\]
and hence Eq. (\ref{G21_SPA}) of the main text. For $\mathbf{R}_{2}%
=\mathbf{\tilde{R}}_{1}\equiv(a,0)$ and hence $\phi(\varphi)=0$, we obtain
$\tilde{g}_{0}=\pi/2$, $\tilde{g}_{y}=2$, $\tilde{g}_{z}=0$. For
$\mathbf{R}_{2}$ close to $\mathbf{\tilde{R}}_{1}$, we use the Jacobi--Anger
expansion $e^{iz\cos\varphi}=\sum_{n=-\infty}^{\infty}i^{n}J_{n}%
(z)e^{in\varphi}$ to obtain%
\begin{align*}
\tilde{g}_{0} &  \equiv\frac{\pi}{2}J_{0}(\Phi)+\sum_{n\geq1}^{\infty}%
i^{n}J_{n}(\Phi)\cos(n\theta)(\xi_{n-1}+\xi_{n+1}),\\
\tilde{g}_{y} &  \equiv2J_{0}(\Phi)+2\sum_{n\geq1}^{\infty}i^{n}J_{n}%
(\Phi)\cos(n\theta)\xi_{|n|},\\
\tilde{g}_{z} &  \equiv i\sum_{n\geq1}^{\infty}i^{n}J_{n}(\Phi)\sin
(n\theta)(\xi_{n-1}-\xi_{n+1}),
\end{align*}
where $J_{m}(x)$ is the $m$th-order Bessel function and
\[
\xi_{\pm n}\equiv\int_{-\pi/2}^{\pi/2}\cos\varphi e^{in\varphi}d\varphi
=\frac{-2\cos(n\pi/2)}{n^{2}-1}.
\]
The value $\xi_{\pm1}=\pi/2$ is obtained by taking the limit $n\rightarrow1$.

\section{Carrier propagators in asymmetric TI P-N junction}

\begin{figure}[t]
\includegraphics[width=\columnwidth]{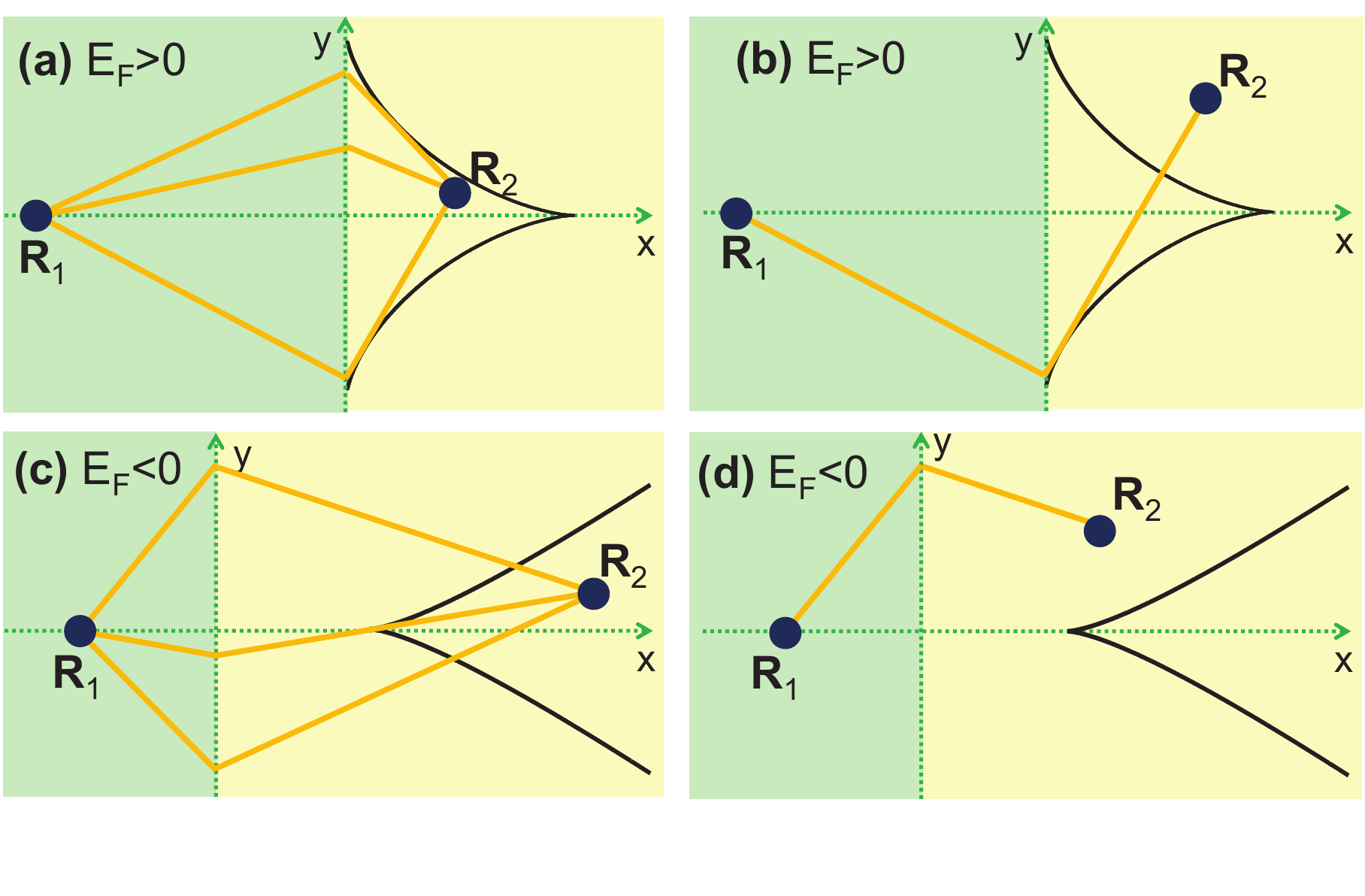}\caption{Classical
trajectory (orange lines) connecting $\mathbf{R}_{1}$ and $\mathbf{R}_{2}$ on
opposite sides of a TI P-N junction. (a) and (b): $E_{F}>0$. (c) and
(d):\ $E_{F}<0$. The black curves denote the caustics.}%
\label{G_PNJ_ASYMMETRIC}%
\end{figure}

We consider a right-going incident electron with energy $E_{F}$ in the N
region with an incident angle $\varphi_{N}$, momentum $(q_{N,x},q_{y})\equiv
q_{N}\mathbf{e}_{\varphi_{N}}$, group velocity $\mathbf{v}_{\mathrm{N}}\equiv
v_{F}\mathbf{e}_{\varphi_{N}}$, and spin state
\[
|\mathbf{e}_{z}\times\mathbf{v}_{\mathrm{N}}\rangle=\frac{1}{\sqrt{2}}%
\begin{bmatrix}
e^{-i\varphi_{N}/2}\\
ie^{i\varphi_{N}/2}%
\end{bmatrix}
,
\]
where $q_{N}\equiv(V_{0}+E_{F})/v_{F}$ is the Fermi momentum in the N region
and $q_{N,x}\equiv(q_{N}^{2}-q_{y}^{2})^{1/2}$. The reflection electron has a
momentum $(-q_{N,x},q_{y})=q_{N}\mathbf{e}_{\pi-\varphi_{N}}$, group velocity
$\mathbf{v}_{\mathrm{r}}=v_{F}\mathbf{e}_{\pi-\varphi_{N}}$, and spin state
\[
|\mathbf{e}_{z}\times\mathbf{v}_{\mathrm{r}}\rangle=\frac{1}{\sqrt{2}}%
\begin{bmatrix}
e^{i\varphi_{N}/2}\\
-ie^{-i\varphi_{N}/2}%
\end{bmatrix}
.
\]
The transmission electron has a momentum $(-q_{P,x},q_{y})=q_{P}%
\mathbf{e}_{\pi-\varphi_{P}}$, group velocity $\mathbf{v}_{\mathrm{P}}\equiv
v_{F}\mathbf{e}_{-\varphi_{P}}$, and spin state
\[
|\mathbf{e}_{z}\times\mathbf{v}_{\mathrm{P}}\rangle=\frac{1}{\sqrt{2}}%
\begin{bmatrix}
e^{i\varphi_{P}/2}\\
ie^{-i\varphi_{P}/2}%
\end{bmatrix}
,
\]
where $q_{P}\equiv(V_{0}-E_{F})/v_{F}$ is the Fermi momentum in the P region
and $q_{P,x}\equiv(q_{P}^{2}-q_{y}^{2})^{1/2}$. From the continuity equation
at the P-N\ interface $x=0$,
\[
|\mathbf{e}_{z}\times\mathbf{v}_{\mathrm{N}}\rangle+r|\mathbf{e}_{z}%
\times\mathbf{v}_{\mathrm{r}}\rangle=t|\mathbf{e}_{z}\times\mathbf{v}%
_{\mathrm{P}}\rangle,
\]
we obtain the transmission amplitude $t(q_{y})=\cos\varphi_{N}/\cos
[(\varphi_{P}-\varphi_{N})/2]$. The continuity of $q_{y}$ across the P-N
interface dictates%
\begin{equation}
\frac{\sin\varphi_{N}}{\sin\varphi_{P}}=\frac{q_{P}}{q_{N}}=\frac{V_{0}-E_{F}%
}{V_{0}+E_{F}}\equiv n. \label{N_DEF}%
\end{equation}

The carrier propagator from $\mathbf{R}_{1}$ to $\mathbf{R}_{2}$ is%

\[
\hat{G}(\mathbf{R}_{2},\mathbf{R}_{1},E_{F})=\int\frac{dq_{y}}{2\pi}%
t(q_{y})e^{i\phi(q_{y})}\frac{|\mathbf{e}_{z}\times\mathbf{v}_{\mathrm{P}%
}\rangle{\langle}\mathbf{e}_{z}\times\mathbf{v}_{\mathrm{N}}{|}}{iv_{F}%
q_{N,x}/q_{N}},
\]
where $\phi(q_{y})=q_{y}Y+aq_{N,x}-q_{P,x}X$. The classical trajectory going
from $\mathbf{R}_{1}$ to $\mathbf{R}_{2}$ is determined by $\partial_{q_{y}%
}\phi=0$ as
\[
Y=a\tan\varphi_{N}-X\tan\varphi_{P},
\]
which, together with Eq. (\ref{N_DEF}), determines the incident angle
$\varphi_{N}$ and the refractive angle $\varphi_{P}$ for the classical
trajectory. For clarity, we denote this classical incident (refractive) angle
by $\theta_{N}$ ($\theta_{P})$. For $n<1$ ($n>1$), there is a unique solution
when $\mathbf{R}_{2}$ lies on the right (left) of the caustics, or three
solutions when $\mathbf{R}_{2}$ lies on the left (right) of the caustics, as
shown schematically in Fig. \ref{G_PNJ_ASYMMETRIC}. The contribution from
different classical trajectories to the propagator are additive. Usually, as
long as $\mathbf{R}_{2}$ is far from the caustics, the propagator is dominated
by the middle trajectory. The contribution from this trajectory as
characterized by $(\theta_{N},\theta_{P})$ is%
\[
G(\mathbf{R}_{2},\mathbf{R}_{1},E_{F})\approx\frac{t}{2\pi iv_{F}\cos
\theta_{N}}|{\mathbf{e}}_{z}\times\mathbf{v}_{\mathrm{P}}\rangle
{\langle\mathbf{e}}_{z}\times{\mathbf{v}}_{\mathrm{N}}{|}\int_{-\infty
}^{\infty}e^{i\phi(q_{y})}dq_{y},
\]
where $t=\cos\theta_{N}/\cos[(\theta_{P}-\theta_{N})/2]$ is the transmission
amplitude of the classical trajectory. The propagation phase along the
classical trajectory (i.e., $q_{y,\mathrm{c}}=q_{N}\sin\theta_{N}=q_{P}%
\sin\theta_{P}$) is
\[
\Phi\equiv\phi(q_{y,\mathrm{c}})=q_{N}\frac{a}{\cos\theta_{N}}-q_{P}\frac
{X}{\cos\theta_{P}}.
\]
Under the stationary phase approximation, we replace $\phi(q_{y})$ by its
second-order Taylor expansion%
\[
\phi(q_{y})\approx\Phi-\frac{1}{2}\frac{\tilde{R}}{q_{F}\cos^{2}\theta_{N}%
}(q_{y}-q_{y,\mathrm{c}})^{2}%
\]
[$\tilde{R}$ is given by Eq. (\ref{RT}) of the main text] to obtain Eq.
(\ref{G_ASYM}) of the main text.


\end{document}